\documentclass[11pt,twoside]{article}

\usepackage{asp2006}
\usepackage{epsf}
\usepackage{psfig}
\usepackage{lscape}

\begin{document}
Revised version of review submitted to the RS Oph meeting, Keele 2007
\title{Mass-loss from Red Giants}   
\author{Brian R.\ Espey \& Cian Crowley}   
\affil{School of Physics, Trinity College Dublin,\\ Dublin 2, Ireland}    

\begin{abstract} 

Although much is known about the nature of winds from hot stars and giants and supergiants with spectral types earlier than K, there is still much to be learned regarding the mass-loss process in cool, late-type stars. We will review the current state of research, with particular reference to observations and modelling of mass-loss from giant stars in symbiotic systems.

\end{abstract}


\section{Introduction}   

When one thinks of mass-loss from cool evolved stars,  the objects that generally spring to mind are heavily-evolved Asymptotic Giant Branch (AGB) stars or planetary nebulae. The cumulative effects of mass-loss over the star's journey up the giant branches are apparent during these spectacular phases of stellar evolution where large amounts of material surround the central object. However, the appreciable amounts of processed material that are lost during the First Red Giant Branch  (FRGB) phase can easily be overlooked when examining the bigger picture of mass-loss from cool, evolved objects due to two main reasons: 1) the effects of these winds are less spectacular and usually only detectable through relatively subtle (compared to AGB objects) examination of wind signatures in the spectra; 2) our lack of understanding of the winds from these earlier stars. The slow, massive winds from heavily evolved AGB stars can be understood in terms of global pulsations lifting gas out to distances above the photosphere where dust can form. Radiation pressure on this cooler material is then very efficient at further driving the dust (and hence the gas) from the star and maintaining the outflow \citep[e.g.][and references within]{2007astro.ph..2444H}. Thus, while the mass-loss theory for cool evolved stars appears relatively successful, the wind acceleration mechanism for less evolved stars still remains unknown.

\subsection{Winds from Symbiotic Red Giants} 

In terms of symbiotic binaries, the existence of a strong wind from the red giant component  is a prerequisite for the symbiotic phenomenon.   Since $\sim$80\% of symbiotics (including RS Oph) are S-type \citep[containing non-dusty, non-heavily pulsating giants,][]{2000A&AS..146..407B} it becomes apparent that a real understanding of symbiotics relies on an understanding of the mass-loss process for early giant-branch stars.
The aim of this review is to examine the present state of understanding of the wind generation, behaviour and characteristics for cool non-dusty giants applicable to S-type symbiotic binaries powered by wind accretion.

\section{Overview of the Mass-loss Problem}

\begin{figure}[h]
\plotone{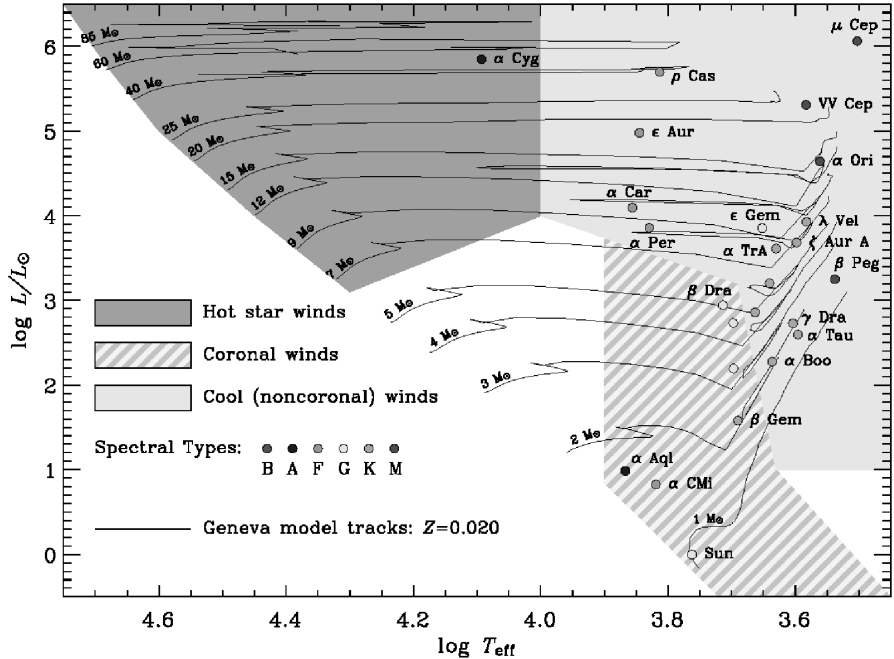}
\caption{HR diagram showing the positions of mass-losing stars from differing regimes.
\label{fig1}}
\end{figure}

Stellar winds are coarsely divided into 3 broad categories (see Figure~\ref{fig1}). Hot winds from luminous OB-type stars are known to be radiatively driven \citep[e.g.][]{1994IAUS..162..475O,1997LNP...497...69L}. Very hot and tenuous coronal winds from cooler main sequence stars and late F and  early K giants are driven by the pressure of the hot gas above the atmosphere (e.g.\ \citet{1999isw..book.....L}). For cooler (K type or later) evolved stars the winds again show still different characteristics.\footnote{For a discussion on the multiple dividing lines between coronal, hybrid and non-coronal cool giants and refinements on these,  see \citet{1979ApJ...229L..27L,1981ApJ...250..293A,1996A&A...310..813R,2003ApJ...598..610A}} 

Since there are few ultraviolet continuum photons to act on the strong atomic resonance transitions in that region of the spectrum, it is not possible to radiatively drive the wind in the same way as for the hot stars. In addition, because cool winds are both more massive and of higher density than coronal winds, cooling processes are also more efficient, thus making corona formation impossible. Since it is not possible to heat the wind to coronal temperatures the wind cannot be pressure driven either. For cool winds from evolved (non-dusty) stars, no known mechanism (nor any combination) can yet satisfactorily explain the observed wind characteristics  \citep[for further discussion, see][]{1991A&ARv...2..249L,1996ASPC..109..481H}. 
The problems for cool star winds stem from a number of requirements to which any successful theory must adhere. Firstly, the relatively low terminal velocities ($<100$ km s$^{-1}$) that are observed from these objects place tight constraints on where the wind generation energy can be deposited (i.e., predominantly in the subsonic region).  Secondly, since the material at the base of the wind can  cool very efficiently through radiative losses, it follows that the energy that drives the wind must be transferred to the outer atmospheric layers through momentum addition rather than through heat transfer. Also, since most of the wind-generation energy is required to lift the material out of the gravitational well of the giant, we actually observe only a small fraction of this energy in the form of the outflow at the terminal velocity, making it difficult to probe the subsonic region at the base of the wind where most of the energy is deposited \citep[e.g.][]{1985ASSL..117..229H}.

\section{Cool Wind Parameters}

Considering their importance to stellar and galactic evolution models, it is unfortunate that the {\bf mass-loss rates} of non-dusty evolved stars are often extremely difficult to measure accurately, with different techniques often resulting in order of magnitude discrepancies. Additionally, there are also many objects with no published estimates of mass-loss. Typical values quoted are $ 10^{-9}<\dot{M}<10^{-5}$, with $\dot{M}$ higher for the more evolved, cooler objects \citep[for further discussion and references within see][]{1998ESASP.413...83L}. 

The {\bf terminal velocities} of these winds are typically measured using the blueward absorption component of the MgII resonance doublet ($\sim$ 2800\AA) which is formed in the expanding wind \citep[see][for examples and discussion]{1987euwi.book..321D}. Typical values for $v_{\infty}$ are $<100$ km s$^{-1}$ with decreased velocities 
observed for cooler objects. We caution, however, that the very low wind speeds derived for the most evolved AGB stars are typically measured using molecular emission lines originating from a circumstellar shell and hence these do not not diagnose the inner wind acceleration region probed by Mg$^+$  in the earlier giants. Therefore, a direct comparison of terminal velocities - and sometimes other wind properties - between these objects may not always be valid.

For non-coronal giants {\bf wind temperatures} in the acceleration region are expected to remain chromospheric (i.e., $<10,000\,$K). This would imply that Mg$^{+}$ is the predominant ionisation stage in the wind. This wind temperature for stars firmly redward of the coronal dividing line is confirmed by results from an analysis of FeII emission features \citep[for a broader discussion and for references within, see][]{2001ASPC..223..368H}, and also the results from our analysis of EG And data \citep[{2006PhDT.........4C,2008ApJ..675..711H}].

In terms of {\bf wind structure and variability},
large variations in mass-loss rates and terminal velocities of both a K supergiant and an early M giant using {\sl IUE} and {\sl HST} data were found by \citet{1998ApJ...495..927M}. 
This finding may be linked to the variable velocity structure observed in absorption profiles diagnosing the wind of cool supergiant stars by \citet{grif} and can be interpreted as clumpy material undergoing episodic ejection from the host star. 

\begin{figure}[th]
\plotone{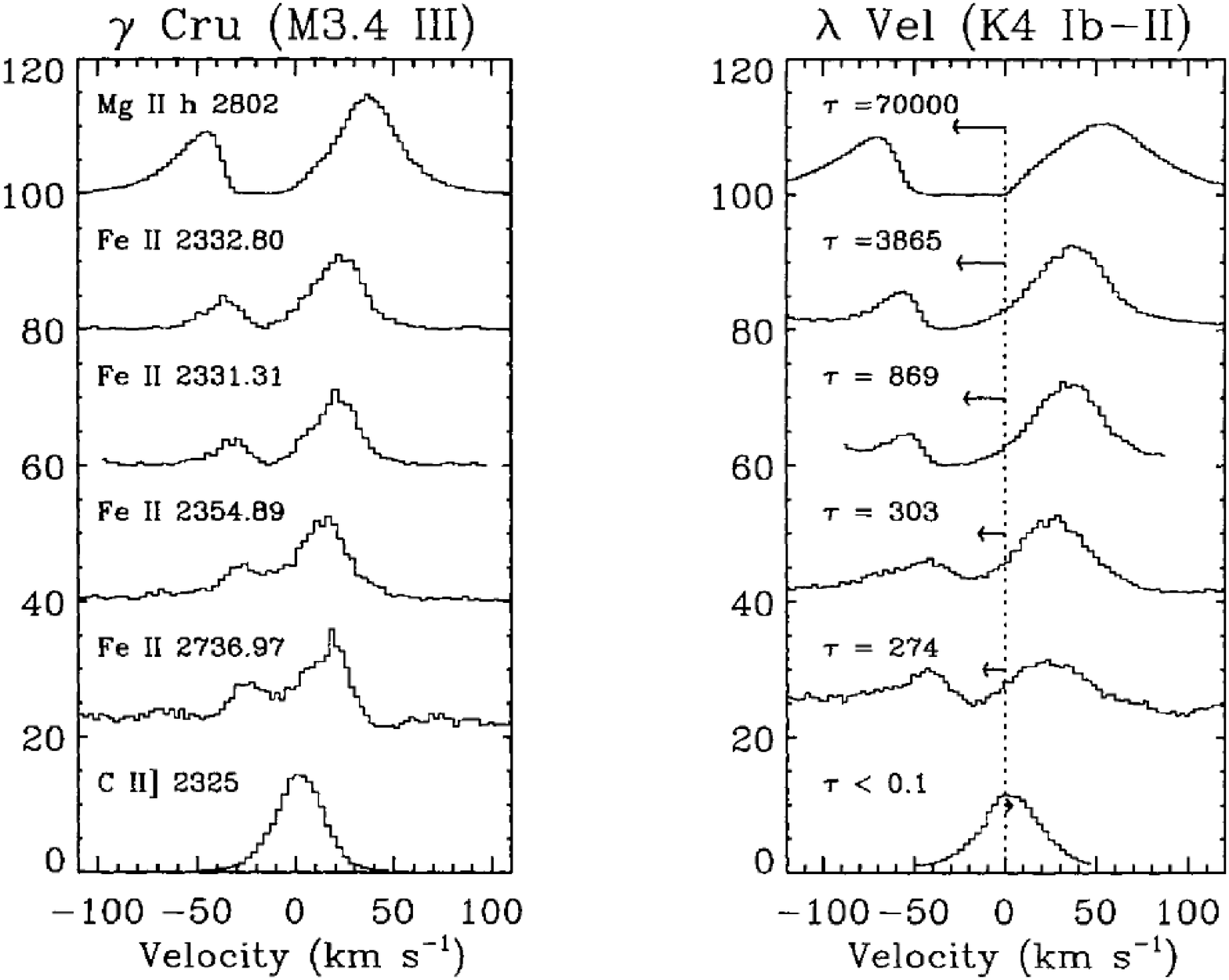}
\caption{It is possible to examine differing layers of the wind in isolated cool stars by examining self-absorbed emission lines of differing optical depths (probing different formation depths). The examples in this figure are taken from \citet{1997ASPC..123...87B}, 
This technique has the advantage that the giant is not affected by a companion, however each profile is disk-averaged and hence its interpretation depends on the the assumed optical depth {\it vs.\/} radius model. Unfortunately, this can result in large uncertainties on any derived parameters. \label{fig2}}
\end{figure}

\section{Symbiotics as Testbeds for Studying Evolved Star Winds}

\subsection*{Radio estimates of mass-loss rate}
\citet{1975MNRAS.170...41W} were the first to point the way to the determination of steady-state mass-loss rates of isolated stars by using infrared or radio measurements of optically thick free-free emission. Subsequently this work was amended and extended by \citet{1984ApJ...284..202S}, and \citet{1992ApJ...387..624S}, to represent steady-state ionised red giant winds, through the introduction of an {\sl 'X'\/} parameter to reflect the proportion and shape of the giant wind photoionised by the white dwarf.
As pointed out by Miko{\l}ajewska and co-workers \citep{2003ASPC..303..478M}, radio emission from the winds of quiescent S-type systems is faint, whereas that from the more luminous outbursting systems is variable, and also complicated by material from previous ejections. Without fully contemporaneous multi-spectral data the application of simplified equations is fraught with possible errors, but Miko{\l}ajewska and collaborators \citep{2003ASPC..303..478M,2002AdSpR..30.2045M,2001MNRAS.324.1023M} have attempted to examine the mass-loss of quiescent S-type binaries using a combination of historical data,
continuum emission estimated from UV and optical data, and also new mm and sub-mm observations. 
In the radio approach, observations of the flux density of optically thick emission, or the turnover from optically thick to thin emission are combined with some assumptions, leading to predicted errors of up to a factor of 2 in the mass-loss rate determination
where there is only partial photoionisation of the giant wind, or even larger errors of up to 2--3 orders of magnitude for cases of Roche lobe mass transfer or a bipolar wind \citep{1993ApJ...410..260S}. Coupled with these uncertainties, we point out that the mass-loss rate
is also dependent on the assumed terminal velocity which - in the case of EG And at least - can be a factor of a few larger than the typically assumed $\sim few \times 10 {\rm ~km~s^{-1}}$.

\subsection*{Direct Measurements of Spatial Structure}
Observations of isolated giants require a combination of both data and assumptions when deriving wind velocity fields from observations of absorbed chromospheric lines; e.g., \citet{1997ASPC..123...87B} - see also Figure~\ref{fig2}. Similarly, estimates of mass-loss in binaries using radio techniques require their own set of assumptions. 
A more direct approach is to make use of binarity, particularly in the case of {\bf eclipsing symbiotic binaries}(see \citet{2005AJ....129.1018H} for a non-symbiotic example). These systems have multiple benefits: 1) they contain a UV/far-UV continuum source; 2) at the same time, they are the widest separation interacting binary systems; 3) the spectra of the two stars are sufficiently different that disentangling the spectral information is straightforward; 4) the changing relative position of the two stars provides multiple sightlines (at least in, or close to, the orbital plane); and, 5) the small relative size of the white dwarf permits accurate column density estimates for a very small region.

\subsection*{Insights from far-UV Observations: EG And}
\begin{figure}[!th]
\plotone{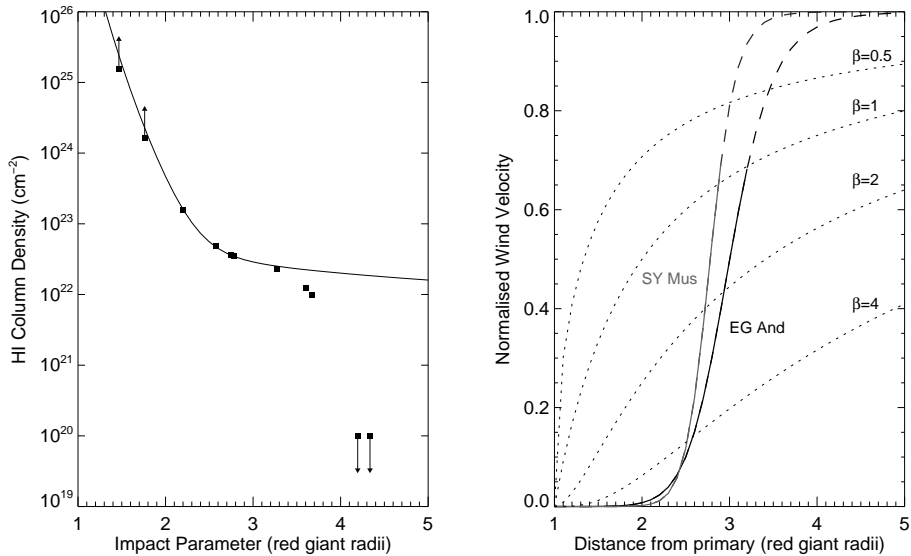}
\caption{{\bf Left:} Column density structure of the red giant wind in EG And derived from STIS and {\sl FUSE} data (error bars typically smaller than point size). Note that the onset of ionisation by the white dwarf occurs beyond $\approx$3 giant radii, exterior to the region of wind acceleration and our temperature estimates. Overlaid is a model of the HI column density falloff.
 {\bf Right:} Plot of the run of the velocity profile of the giant derived from the model shown on the left using the inversion technique described by \citet{1993A&A...274.1002K}. Also plotted is the wind profile derived for SY Mus (with {\sl IUE}) data using the same technique. The deviation of the velocity profile from the $\beta$ laws, which are commonly used to parametrise the wind profile in late giants, is striking. Profile sections that are dashed are extrapolated from the fit to the column densities below $\approx 3~R_{RG}$\ since the wind becomes ionised at these higher impact parameters.  \label{fig3}}
\end{figure}

Our best-studied eclipsing symbiotic object is EG Andromedae, for which we have multiple epoch ultraviolet  {\it HST}/STIS and {\it FUSE} data \citep[see][]{2006PhDT.........4C,2008ApJ..675..711H}.
Note that in terms of both the giant's spectral type (M2.4) and binary separation this system is similar to RS Oph although EG And has never been seen in outburst, and this is corroborated by the low luminosity of the white dwarf ($<$30 $L_{\sun}$). Initial work on {\it IUE} data by \citet{1991A&A...249..173V} showed that the velocity field of the red giant does not conform to the expectations of $\beta$-law models, with a delayed wind acceleration (see Figure~\ref{fig3} for a velocity law derived from our data).

\subsection*{Results}
Our main result is that our newer and higher signal-to-noise data confirm Vogel's earlier finding of a wind that deviates significantly from the $\beta$-law assumption, with an acceleration setting in at $\sim2$\ giant radii. Our UV/far-UV observations also enable us to determine the temperature of the wind, and it is consistent with an isothermal medium of $\sim8,000\,$K over $\leq3 {R_{RG}}$\ which is much hotter than the giant's effective temperature. Note that this temperature is determined for wind material that is {\it not} subject to the ionising effects of the white dwarf (see Figure~\ref{fig3} - left), so is representative of the true conditions.

\begin{figure}[!th]
\plotone{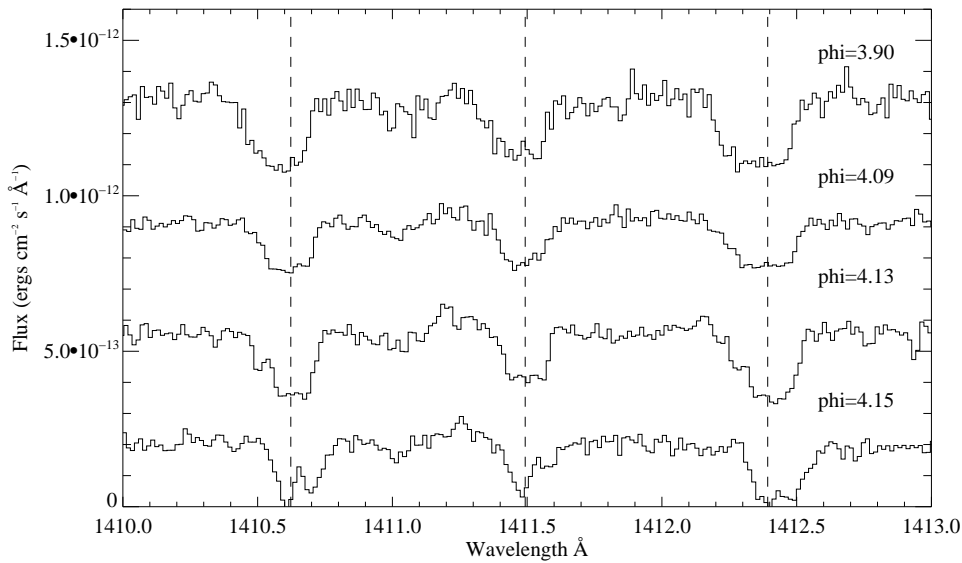}
\caption{Variations in the absorption lines wind line profiles from STIS (dashed lines correspond to the line centres at the EG And systemic velocity). Note the structure that 
is resolved in the profiles implying a certain clumpiness in the inner portion of the outflow. \label{fig5}}
\end{figure}

Our {\sl HST}/STIS observations further suggest that clumping occurs in the wind, as lines can be resolved into multiple absorption components (see Figure~\ref{fig5}) with velocity widths  of $\sim$3--5 km s$^{-1}$. Whether the components are evidence for large-scale clumping or a more flocculent structure is not determined, but the size scale of the clumps is similar to the isothermal scale height for the material.
Despite the possible complications of this structure, and the possibility of some enhancement of mass loss in the plane of the binary, our estimated mass-loss rate is reasonably consistent with the estimate of \citet{1993ApJ...410..260S}, based on radio observations of the circumbinary nebula.

We have also set a stringent limit on the presence of molecular hydrogen in the wind of N(H$_2$)/N(HI) $< 10^{–8}$, confirming the unimportance of molecular driving in early M stars.
From the Mg{\sc II} resonance profile 
we derive a relatively high terminal velocity for this wind of 75 km s$^{-1}$ - much larger than is typically assumed. Equating this velocity with $v_{esc}$\ at wind origin results in a radius of $\approx1.4 {R_{RG}}$\ - just inside the acceleration zone (see Figure~\ref{fig3}, right panel).
 
\begin{figure}[!th]
\plotone{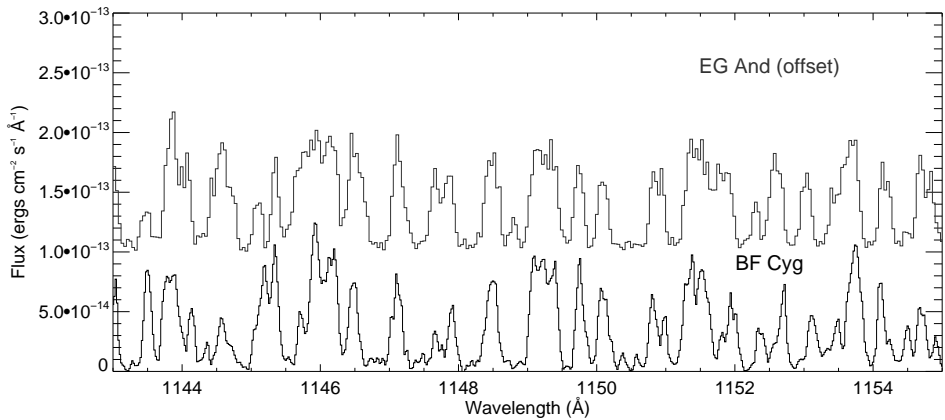}
\caption{Comparison of wind occulted {\sl FUSE} data for EG And (top) and BF Cyg (bottom). Both observations were taken at similar orbital phases and diagnose the cool wind of the giant star in absorption through a host of narrow lines. The lines is this region of spectrum are primarily {Fe\sc II}, {P\sc II}, {N\sc I} and {S\sc II}. The similarities between the wind spectra are striking, especially considering that both giants are of differing spectral type and mass loss. The data were corrected for radial velocity differences and the BF Cyg data were scaled  to compensate for the differences in far-UV brightness for the two objects. \label{bfplot}}
\end{figure}

Preliminary comparisons with another of our targets (BF Cyg; M5III), shows an overall similarity with the EG And absorption data (see Figure~\ref{bfplot}), although with $\sim 20$\ times the mass-loss rate BF Cyg has larger column densities. The close agreement of the spectra suggests that the more complete and detailed EG And data may serve as a Rosetta stone in the analysis of other systems.

\section{The Wind Driving Mechanism}
\citet{1965SSRv....4..666P} first pointed out that magnetically driven mass motions along open field lines could explain the non-steady component of the solar wind. This Alfv\'{e}n mechanism has also been put forward as the explanation for stellar winds across the HR diagram in cases where neither radiative nor pressure-driven mechanisms are possible. Alfv\'{e}n waves can solve the problem of generating a sufficiently strong wind, but only if the energy input can be damped low in the atmosphere in order to bring the terminal velocity of the wind into line with observed values, as discussed by \citep{1980ApJ...242..260H}.

A recent paper by \citet{2007ApJ...659.1592S} provides a convincing demonstration of the progress that has been made in the theoretical modelling of outflows. Suzuki addresses the problem in a self-consistent manner whereby the wind is a natural consequence of energy input from the photosphere. Using results from solar work as a guide where necessary, to model wind driving for stars which straddle the coronal dividing line in the HR diagram (right-hand edge of coronal region in Figure~\ref{fig1}. On the hot side of the coronal line efficient acceleration can occur in a time-variable two-phase medium, but on the cool side, the denser wind suppresses the coronal component.
His model also nicely explains the delay in onset of the acceleration region as a consequence of initial lateral expansion of the magnetised bubble, followed by subsequent rapid radial expansion above 
$\sim2 {R_{RG}}$, as is observed in the case of the  acceleration regions shown in Figure~\ref{fig3}.

\section{Applicability of Symbiotic Results to the General Stellar Population?}
We have noted the quiescent nature of the EG And system, and its usefulness as a proxy for mass loss from isolated field stars. Perhaps the strongest support for this finding is that in the optical - aside from relatively weak emission lines - the giant spectrum bears a striking similarity of the optical spectrum to that of an isolated thick disk spectral standard.

Our absorption results alone can not address the question of symbiotic wind (a)symmetry, examining as they do material close to the plane of the system through their line-of-sight velocities. It is to be expected that the companion will simultaneously both focus and disturb the mass flow to some extent.
However, in mitigation, it should be pointed out that the region we have studied lies {\it interior\/} to the binary orbit in a region which is expected to be relatively undisturbed, as shown graphically in the simulations of models with three very different ratios of red giant/white dwarf wind momenta in \citet[]{2000ASPC..204..331W}.

In terms of the gravitational influence of the dwarf, the observed symmetry in the wind line profiles in each STIS spectrum of EG And shows that the wind material is not being stripped from the giant in any observed sightline direction (i.e., in the orbital plane), and in terms of both mass-loss rate and mean density, radio data are consistent with UV results. It thus appears that conditions derived from the absorbing  material are intrinsic to the red giant chromosphere/lower wind, and hence are not dramatically altered by the hot component. More generally, however, these results will vary from system to system and hence require further investigation. 

\section{Conclusions}
Through the eras of {\sl IUE}, {\sl HST} and {\sl FUSE}, much has been learnt about cool giant winds and circumstellar material. However, this information has also highlighted the  complexity involved in modelling and understanding the underlying processes. While progress has been steady over the past three decades, a full understanding of RGB mass-loss is still some way off. In the short term the most significant progress can be made through the development of models of greater complexity, particularly since the recent demise of all instruments with the capacity for high-resolution UV/far-UV spectroscopy. 

However, there is also hope on the observational side due to the planned refurbishment of {\sl HST} with COS and (potentially) a repaired STIS. A STIS survey of additional eclipsing symbiotic systems would provide sufficient observations at a high enough spectral resolution to analyse velocity structure in both self-absorbed diagnostic lines (Mg{\sc II} h \& k, Fe{\sc II} etc.), as well as the narrow absorption lines seen in the white dwarf continuum. 
Due to its better sensitivity, access to COS will permit the extension of the sample of objects with well-determined wind velocities (from the isolated Mg{\sc II} profiles) in a reasonable amount of telescope time, albeit without the additional spectroscopic information on clumpiness available via.\ STIS.
A detailed comparison of the column density measurements, radio results, and self-absorption results for a larger, well-determined sample of symbiotic giants would provide the clues necessary.

\acknowledgements 

The authors wish to thank our colleagues for numerous insightful discussions as well as P.\ Bennett for kindly providing Figure 1. Support for this work was provided by Enterprise Ireland Basic Research 
grant SC/2002/370 from the EU funded NDP, and Science Foundation Ireland grant 06/RFP/PHY010.
{\sl FUSE} data were obtained under the Guest Investigator Program and  and
supported
by NASA grants NAG5-8994 and NAG5-10403 to the Johns Hopkins University
(JHU).
STIS data were obtained under the Guest Observer program STSI-GO0948701.



\end{document}